\documentclass[preprint,showpacs,preprintnumbers,amsmath,amssymb]{revtex4}

\pdfoutput=1
\usepackage{graphicx}\graphicspath{./graph/}%
\usepackage{dcolumn}% Align table columns on decimal point
\usepackage{bm}% bold math
\usepackage{color}
\usepackage{verbatim}
\usepackage{amssymb}
\usepackage{amsfonts}
\usepackage{latexsym}
\usepackage{epstopdf}

\definecolor{red}{rgb}{1,0,0}
\definecolor{green}{rgb}{0,1,0}
\definecolor{blue}{rgb}{0,0,1}

\newcommand{\be}{\begin{equation}}
\newcommand{\ee}{\end{equation}}
\newcommand{\bea}{\begin{eqnarray}}
\newcommand{\eea}{\end{eqnarray}}
\newcommand{\bdm}{\begin{displaymath}}
\newcommand{\edm}{\end{displaymath}}

\newcommand\ptl{\partial}
\newcommand\dpr{^{\prime\prime}}

\newcommand\pr{^\prime}

\newcommand\vph{\varphi}
\begin{document}
%\preprint{APS/123-QED}
\title{Improved Numerical Method for Solution of \\ the Ha\" issinski Equation }
\author{Robert Warnock }
\email{warnock@slac.stanford.edu}
\affiliation{SLAC National Accelerator Laboratory, Stanford University, Menlo Park, CA 94025, USA}
\affiliation{Department of Mathematics and Statistics, University of New Mexico, Albuquerque, NM 87131, USA}
\author{Karl Bane }
\email{kbane@slac.stanford.edu}
\affiliation{SLAC National Accelerator Laboratory, Stanford University, Menlo Park, CA 94025, USA}
\begin{abstract}
The longitudinal charge density of an electron beam in its equilibrium state is given by the solution of the
Ha\"issinski equation, which provides a stationary solution of the  Vlasov-Fokker-Planck equation.
The physical input is the longitudinal wake potential.
We formulate the Ha\" issinski equation as a nonlinear integral equation with the normalization
integral stated as a functional of the solution. This equation can be solved in a simple way
by the matrix version of Newtons's iteration, beginning with the Gaussian as a first guess.
We illustrate for several quasi-realistic wake potentials.
Convergence is extremely robust, even at currents much higher than nominal for the storage rings considered.
The method overcomes limitations of earlier procedures, and provides the convenience of automatic
normalization of the solution.

\end{abstract}
\maketitle
\section{Introduction}
The collective longitudinal motion of electrons or positrons in a storage ring seems to be well described by the Vlasov-Fokker-Planck (VFP) equation,
in which the collective force is described by a wake potential which accounts for the electromagnetic environment due to the vacuum chamber.
The equation has solutions that are stationary in time, which may or may not be stable under perturbations, depending on the value of the
beam current. These are solutions of the Ha\" issinski equation \cite{hais}, which may be stated as a nonlinear integral equation or integro-differential equation. To determine the
threshold in current for an instability to appear, one can linearize the Vlasov equation about the Ha\" issinski solution. Alternatively, one
can integrate the VFP equation as an initial value problem in time, with the Ha\" issinski equilibrium as the initial value \cite{bobjim}. In either
approach, computation of the Ha\" issinski solution is an essential first step in determining the threshold for an  instability. (Admittedly, one can also get an idea of stability by running a VFP integration from an arbitrary
initial value, say a Gaussian, but that will lead to a somewhat ambiguous definition of the threshold.)

The method of solution presented here was worked out by the first-named author twenty years ago, but was not published
except for a description in words in Ref.\cite{bobjim}. Although the method was adopted by a few colleagues, it has not become a
standard tool. Since it is quite simple and avoids limitations of other methods, a belated publication  seems worthwhile.

The idea of the method will seem obvious to anyone acquainted with ideas of functional analysis \cite{berger} and their application in
numerical methods \cite{collatz}.
The integral equation for the charge density $\lambda$ is viewed as an equation $F(\lambda)=0$ on an appropriate function space.
The equation is discretized by a numerical quadrature rule for the integrals involved, and then solved by the matrix version of
Newton's method. An essential step is to define $F$ so that a solution is automatically normalized. We shall not be
concerned with a rigorous basis for discretization, but methods to treat that issue are available \cite{anselone}.

What was not so obvious before implementation is the extremely robust convergence of the Newton iterates.  For realistic wake potentials we have never seen
a failure of convergence to machine precision in a few iterations, even at currents far beyond the threshold of instability. Here the starting point
for the Newton iteration was merely a Gaussian, the zero current solution.

In this paper we include results for high current, in order to explore the mathematical properties of the equation, and
to demonstrate the excellent convergence of the iteration. In many physical problems governed by nonlinear equations one finds critical values
of some  strength parameter, for instance in the buckling of a column at a certain value of the load \cite{keller}, (\cite{berger}\ , Chap.4).  As a function of the parameter a solution may branch into two or more solutions,
or become complex, or simply cease to exist in one way or another. It is then natural to look for critical points in the
current parameter in the Ha\" issinski equation.  For the wake potentials considered here we find no such
  points up to very high currents. In fact, we can argue that our solutions are locally unique in the function space
 considered, since bifurcation can occur only at a singularity of the Jacobian of the system \cite{berger}. Our
  Jacobian is always far from singularity. The high current solutions represent unstable equilibria, and will not be realized in the laboratory.

\section{Solution of the Vlasov-Fokker-Planck  Equation for the Equilibrium State \label{equil}}
We are concerned with longitudinal motion within a single bunch of particles in an electron storage ring.
The linearized motion without collective effects is described in terms of the slip factor $\eta$, the dimensionless constant which relates the first order change in revolution frequency $\omega_r$ to a change in momentum $P$:
\be
\eta= -\frac{P_0}{\omega_0}\bigg(\frac{d\omega_r}{dP}\bigg)_{P_0}=\alpha-\frac{1}{\gamma_0^2}\ . \label{etadef}
\ee
Here $\omega_0$ and $P_0$ are the nominal values of revolution frequency and momentum (those for a particle synchronizing with the RF),  $\gamma_0$ is the nominal Lorentz factor, and $\alpha$ is the momentum
compaction factor. (Some authors define $\eta$ with the opposite sign, and some call $\eta$ the momentum compaction factor.)

The dynamical variables of longitudinal motion are often taken to be $\Delta\phi$ and $\Delta E$, where $\Delta\phi$ is the deviation of the  RF phase from its synchronous value at the time the particle encounters the RF field, and $\Delta E$ is the deviation of the energy from the nominal value $E_0$ \cite{sylee}. We prefer to work with equivalent
 dimensionless variables $q$ and $p$, normalized to be of order 1, and a corresponding dimensionless  $\tau$, equivalent to the time \cite{oide}.  We define
\be
  q=\frac{z}{\sigma_z}\ ,\qquad p=-{\rm sgn}(\eta)\frac{E-E_0}{\sigma_E}\ ,\qquad \tau=\omega_s t\ .\label{qpdef}
\ee
Here $z=s-s_0=s-\beta_0ct$ is the distance (in arc length $s$ on the reference orbit) to the synchronous particle at $s=s_0$, thus positive for a leading particle, and ${\rm sgn}(\eta)$ is
the signum function, equal to $1$ for $\eta>0$ and $-1$ for $\eta<0$. The constant $\omega_s=2\pi f_s$ is the circular synchrotron frequency. At first we think of $\sigma_z$ and $\sigma_E$ as  some positive constants to render $q$ and $p$
dimensionless and of order $1$, leaving to later a specific choice of their values. One can  show that $\Delta\phi=-hz/R$, where $h$ is the harmonic number and $R=C/2\pi$, where $C$ is the circumference of the reference orbit followed by the synchronous particle. From this we can write the differential equations \cite{sylee} (which
approximate a discrete map) in terms of the
new variables as follows,
\be
\frac{dq}{d\tau}=\frac{p}{a}\ ,\qquad \frac{dp}{d\tau}=-aq\ ,\qquad a=\frac{\beta_0\omega_s\sigma_z}{c}\frac{E_0}{|\eta|\sigma_E}\ . \label{qpeom}
\ee
The corresponding Hamiltonian is
\be
H(q,p)=a\ \frac{q^2}{2}+\frac{1}{a}\ \frac{p^2}{2}\ . \label{ham}
\ee

In a storage ring with normal equilibration from synchrotron radiation, in which effects of diffusion balance effects of dissipation, the phase space density in the limit of small beam current is
\be
f_0(q,p)=A\exp(-H(q,p))\ ,
\ee
where $A$ is a constant for normalization.
Hence we can interpret $\sigma_z$ and $\sigma_E$ as the r.m.s. bunch length and energy spread for weak current, provided that $a=1$ or
\be
\frac{\beta_0\omega_s\sigma_z}{c}=\frac{|\eta|\sigma_E}{E_0}\ . \label{aeq1}
\ee
Henceforth we choose $\sigma_z$ and $\sigma_E$ to satisfy (\ref{aeq1}), whatever their interpretation.

The probability density in phase space, normalized to 1, is denoted by $f(q,p,\tau)$, and the spatial probability density by $\lambda(q,\tau)$, thus
\be
 \int_{-\infty}^\infty\int_{-\infty}^\infty f(q,p,\tau)dqdp=1\ ,\qquad \lambda(q,\tau)=\int_{-\infty}^\infty f(q,p,\tau)dp\ .           \label{norm1}
\ee
The Vlasov-Fokker-Planck (VFP) equation to determine $f$ is
\be
\frac{\ptl f}{\ptl \tau}+\frac{dq}{d\tau}\frac{\ptl f}{\ptl q}+\frac{dp}{d\tau}\frac{\ptl f}{\ptl p}=\frac{2}{\omega_s t_d}
\frac{\ptl}{\ptl p}\bigg[p\frac{\ptl f}{\ptl p}+\frac{\ptl f}{\ptl p}\bigg]\ , \label{vlas}
\ee
where $t_d$ is the longitudinal damping time.

The single-particle equations of motion, modified to include the Vlasov collective force, are
\be
\frac{dq}{d\tau}=p\ ,\qquad \frac{dp}{d\tau}=-q-F(q,f(\cdot,\tau))\ , \label{singpart}
\ee
where $-q$ is the linear force from RF and $-F$ is the collective force.  Here $F$ is a functional of the distribution $f$ which is assumed to have the form
\be
F(q,f(\cdot,\tau))=I\int_{-\infty}^\infty W(q-q\pr)\lambda(q\pr,\tau)dq\pr= I\int_{-\infty}^\infty W(q-q\pr)\bigg[\int_{-\infty}^\infty f(q\pr,p,\tau)dp\bigg]dq\pr\ .
\label{bigF}
\ee
The wake potential $W$ is defined to have the dimension of a potential per unit charge, and to be positive where it causes an energy gain. It follows that the normalized
current $I$ has the form
\be
I=\frac{{\rm sgn}(\eta)e^2N}{2\pi\nu_s\sigma_E}\ ,        \label{Idef}
\ee
where $N$ is the number of particles and  $\nu_s=\omega_s/\omega_r$ is the synchrotron tune.
The notation $F(q,f(\cdot,\tau))$ is intended to indicate that $F$ depends on all values of $f$ over phase space at time $\tau$. In
some models the wake potential  is zero in front of the bunch ($q>0$) but that is not assumed in the following.

Although the
formula (\ref{bigF}) usually goes unquestioned, it is in fact not the most general form of the collective force for a time dependent charge density.
For a bunch on a curved orbit it does not account for the charge density being different at the retarded time from what it is at the current time \cite{WRVE, WV}.

In view of (\ref{singpart}) the VFP equation takes the form
 \be
\frac{\ptl f}{\ptl \tau}+p\frac{\ptl f}{\ptl q}-\big[q+F(q,f(\cdot,\tau))\big]\frac{\ptl f}{\ptl p}=2\beta
\frac{\ptl}{\ptl p}\bigg[p\frac{\ptl f}{\ptl p}+\frac{\ptl f}{\ptl p}\bigg]\ , \label{vlast}
\ee
where $\beta=(\omega_s t_d)^{-1}$, with $t_d$ being the longitudinal damping time.
The Fokker-Planck terms on the right hand side account for damping and diffusion due to incoherent synchrotron radiation.

We are interested in an equilibrium, a time-independent solution of (\ref{vlast}), denoted by $f_0(q,p)$. We seek such a solution in the Maxwell-Boltzmann form
\be
f_0(q,p)=\frac{1}{\sqrt{2\pi}}\exp(-p^2/2)\lambda(q)\ ,\qquad \int_{-\infty}^\infty\lambda(q)dq=1\ .                   \label{mb}
\ee
The Fokker-Planck terms add up to zero for this Gaussian function of $p$, owing to compensation of diffusion by damping. Thus $f_0$ will
be an equilibrium solution provided that the spatial density $\lambda$ satisfies
\be
\frac{d\lambda}{ dq} +\big[q+F(q,f_0(\cdot))\big]\lambda=0\ , \qquad  F(q,f_0(\cdot))=I\int_{-\infty}^\infty W(q-q\pr)\lambda(q\pr)dq\pr\ .    \label{haisdiff}
\ee
Any solution of (\ref{haisdiff}) may be represented as follows:
\be
\lambda(q)=A\exp\big(-V(q,\lambda(\cdot))\ \big)\ , \qquad   V(q,\lambda(\cdot))=\frac{q^2}{2}-I\int_q^\infty dq\pr\int_{-\infty}^\infty  W(q\pr-q\dpr)
\lambda(q\dpr)dq\dpr\ ,
\label{haisint}
\ee
where the constant $A$ is chosen to enforce the normalization of (\ref{mb}). This follows from separation of variables ( $d\lambda/\lambda=-(q+F)dq$\ ) and integration.
Now it is convenient to reverse the order of integrations, after introducing  the integrated wake potential $S$, where
\be
S(q-q\dpr)=\int_q^\infty W(q\pr-q\dpr)dq\pr=\int_{q-q\dpr}^\infty W(r)dr\ , \label{sdef}
\ee
thus
\be
V(q,\lambda(\cdot))=\frac{q^2}{2}-I\int_{-\infty}^\infty  S(q-q\pr)\lambda(q\pr)dq\pr\ . \label{vwiths}
\ee
The kernels $W(q-q\pr)$ and $S(q-q\pr)$ may be viewed as giving the response to a delta function source and a step function source, respectively \cite{bane1}.

It follows from (\ref{haisint}) and (\ref{vwiths}) that a normalized solution of (\ref{haisdiff}) must satisfy
\be
\lambda(q)=\frac{\exp\big[-q^2/2+I\int S(q-q\pr)\lambda(q\pr)dq\pr\big]}
{\int \exp\big[-{q\pr}^2/2+I\int S(q\pr-q\dpr)\lambda(q\dpr)dq\dpr\big]dq\pr}   \label{ie}
\ee
This nonlinear integral equation (\ref{ie}) is our main object of study. It is convenient to rewrite it as
\be
F(\varphi,I)=0\ ,            \label{Feq0}
\ee
where $ \varphi(q)=I\lambda(q)$ and
\bea
 &&F(\varphi,I)=\nonumber\\
 &&\varphi(q)\int \exp\bigg[-{q\pr}^2/2+\int S(q\pr-q\dpr)\varphi(q\dpr)dq\dpr\bigg]dq\pr-I\exp\bigg[-q^2/2+\int S(q-q\pr)\varphi(q\pr)dq\pr\bigg]\ ,\nonumber \\
  \label{Fdef}
\eea
with all integrations on $(-\infty,\infty)$.
\section{Previous methods of solving the Ha\" issinski Equation}
To motivate our method we briefly review techniques in common use, and point out  limitations that are avoided by our algorithm.
\subsection{Solution by Simple Iteration}
An obvious approach is to generate a sequence  of functions $\{ \lambda^{(0)},  \lambda^{(1)}, \cdots\}$ by the rule
\be
\lambda^{(k+1)}(q)=A \exp\bigg[-q^2/2+\int S(q-q\pr)\lambda^{(k)}(q\pr)dq\pr\bigg]\ ,    \label{plainiter}
\ee
where $\lambda^{(0)}$ is the normalized Gaussian and $A$ has some trial value, say $1/\sqrt{2\pi}$. If the sequence
converges, try again with different values of $A$, searching for a value of $A$ such that the final iterate is normalized to adequate precision. This
could be made more convenient by normalizing every iterate; in other words, just apply simple
iteration to our equation (\ref{ie}) with embedded normalization, so that
\be
\lambda^{(k+1)}(q)=\frac{\exp\big[-q^2/2+I\int S(q-q\pr)\lambda^{(k)}(q\pr)dq\pr\big]}
{\int \exp\big[-{q\pr}^2/2+I\int S(q\pr-q\dpr)\lambda^{(k)}(q\dpr)dq\dpr\big]dq\pr}    \label{iterie})
\ee

 Unfortunately, in numerical experience this sequence or (\ref{plainiter})  fails to converge at larger $I$, including values of practical interest. Rather,
the iterates eventually oscillate between one pattern and another.
This failure has no physical significance, as is shown by successful continuation of the solution to large $I$ by other methods, for instance the one we advocate.
\subsection{Solution of the Equation in Integro-Differential Form}
This method aims to solve the Ha\"issinski equation expressed as the integro-differential equation of (\ref{haisdiff}). This can be done
in a simple way only if $W(q)=0$ for $q>0$, a condition that is not strictly true for numerically determined wake potentials for real storage rings.
In fact such potentials are non-zero in a small region $0<q<a$. A more serious violation of the condition can occur in the case
of coherent synchrotron radiation. Depending on circumstances, it may happen that  $W(q)$ will be non-zero over a large range of positive $q$

We seek a numerical solution which is strictly Gaussian for $q\ge\kappa$, approximating the actual solution which is asymptotic to a Gaussian.
We write $\lambda(q)=A\exp(-q^2/2),\  q\ge\kappa$. Then with
the above mentioned restriction on $W$ the integro-differential equation to solve is
\be
\frac{d\lambda}{dq}= -\bigg[q+\int_q^\infty W(q-q\pr)\lambda(q\pr)dq\pr\bigg]\lambda(q)\ .
\ee
The idea is to start at $q=\kappa$, where the right hand side is known, then integrate backwards in $2N$ steps of
 $-\Delta q=-\kappa/N$ to $q=-\kappa$. If we apply Euler's method, the first two integration steps are as follows:
 \bea
 &&\frac{\lambda(\kappa)-\lambda(\kappa-\Delta q)}{\Delta q}=-\bigg[\kappa+I\int_\kappa^\infty W(\kappa-q)\lambda(q)dq\bigg]\lambda(\kappa) ,\label{step1}\\
 &&\frac{\lambda(\kappa-\Delta q)-\lambda(\kappa-2\Delta q)}{\Delta q}= -\bigg[\kappa-\Delta q+I\int_\kappa^\infty W(\kappa-\Delta q-q)\lambda(q)dq\bigg]
 \lambda(\kappa-\Delta q)\nonumber\\
 && -I\lambda(\kappa-\Delta q)\int_{\kappa-\Delta q}^\kappa W(\kappa-\Delta q-q)\lambda(q)dq\ .\label{step2}
 \eea
The integral in the last term in (\ref{step2}) can be approximated by the trapezoidal rule as
\be
\frac{\Delta q}{2}\big[W(0)\lambda(\kappa-\Delta q)+W(-\Delta q)\lambda(\kappa)\big]\ . \label{trap}
\ee
Thus $\lambda(\kappa-\Delta q)$ and $\lambda(\kappa-2\Delta q)$  are determined by (\ref{step1}), (\ref{step2}) and (\ref{trap}).
Continuing in a similar way we build up the discretized solution $\lambda(\kappa-i\Delta q),\ i=0,\cdots, 2N$, which  depends on the constant $A$ in the initial condition.
The process must be repeated to search for an $A$ such that the solution is normalized.

The solution, unnormalized in general, is well-defined for any current $I$, so if normalization can be achieved we
have overcome the restriction to small current required in the iterative method. Unfortunately we see no simple way to automate the
normalization. The awkwardness in normalization, and the requirement that $W(q)$ vanish for $q>0$, are two
undesirable features of this method that we wish to avoid.
\subsection{Solution by Time-Domain Integration of the Vlasov-Fokker-Planck Equation}
Another possibility is to integrate the full VFP equation (\ref{vlast}) as an initial-value problem, using the
method of local characteristics \cite{bobjim}. With $f(q,p,0)=\exp(q^2+p^2)/2\pi$ as the initial value, the solution is expected to converge
to the Ha\" issinski solution at large $\tau$, provided that the current is below the threshold for instability. The disadvantage
of this approach is that is does not allow the study of currents above threshold, and it takes much more computer time. It does
provide, however, a useful check of the VFP solution algorithm, given a Ha\"issinski solution from another method.

\section{Numerical Solution of the Nonlinear Integral Equation by Newton's Method \label{newton}}
We discretize the equation (\ref{Feq0}) on a uniform mesh of $n$ points $q_i$, running from $-\kappa$ to $\kappa$:
\be
q_i=-\kappa+(i-1)\Delta q\ ,\quad \Delta q=\kappa/m\ , \quad i=1,2,\cdots, n=2m+1\ .   \label{mesh}
\ee
We write $\vph_i$ for the numerical approximation to $\vph(q_i)$, and $S_{i-j}$ for $S(q_i-q_j)$. We discretize the
integrals by some quadrature rule with weights $w_i$. Then the discretized form of (\ref{Feq0}) is
\bea
&&F_i(\vph,I)=\vph_i\sum_jw_j\exp\big[-q_j^2/2+\sum_kw_kS_{j-k}\vph_k\big]-I \exp\big[-q_i^2/2+\sum_jw_jS_{i-j}\vph_j\big]=0\ ,\nonumber\\
&&\hskip 2cm i=1,\cdots,n\ .   \label{Feq0dis}
\eea

Newton's method defines a sequence of approximations by successive linearizations of the equation. If $\vph^{(p)}$ is the $p$-th approximate solution, then
$\vph^{(p+1)}$ is obtained from the first order Taylor development  about $\vph^{(p)}$:
\be
F_i( \vph^{(p)},I)+\sum_j\frac{\ptl F_i( \vph^{(p)},I)}{\ptl\vph_j}(\vph_j^{(p+1)}-\vph_j^{(p)})=0\ , \quad i=1,\cdots,n\ . \label{iter}
\ee
An initial guess $\vph^{(0)}$, sufficiently close to the desired solution, is required. The Jacobian matrix element  computed
 from (\ref{Feq0dis}) is
\bea
&&\frac{\ptl F_i( \vph,I)}{\ptl\vph_j}=\sum_kw_k\big(\delta_{ij}+\vph_iw_jS_{k-j}\big)\exp\big[-q_k^2/2+\sum_lw_lS_{k-l}\vph_l\big]\nonumber\\
&&\hskip 2.3cm -Iw_jS_{i-j}\exp\big[-q_i^2/2+\sum_kw_kS_{i-k}\vph_k\big]\ . \label{jacobian}
\eea
Given $\vph^{(p)}$, we compute $F(\vph^{(p)},I)$ and $\ptl F(\vph^{(p)},I)/\ptl\vph$ from (\ref{Feq0dis}) and (\ref{jacobian}) and then
solve the system (\ref{iter}) of $n$ linear equations for $x=\vph^{(p+1)}-\vph^{(p)}$ to find the update $\vph^{(p+1)}=x+\vph^{(p)}$.

A convenient criterion for convergence may be stated in terms of a vector norm, for instance
\be
\| \vph \|=\sum_{i=1}^n |\vph_i|\ .         \label{norm}
\ee
We judge convergence by the quantity
\be
r=\frac{\| \vph^{(p+1)} -\vph^{(p)}\|}{\| \vph^{(p)}\|}\ ,   \label{rdef}
\ee
demanding that it reach a small value, say $10^{-14}$ , as $p$ increases. Of course, one must also check convergence under refinement of the mesh (\ref{mesh}). We
normally do that just by graphical comparisons, but it could be done more quantitatively.

At sufficiently small current the Gaussian should be a suitable first guess, $\vph^{(0)}=I\exp(-q^2/2)/\sqrt{2\pi}$. In practice this choice is  good for realistic currents with reasonable wake potentials, in fact at currents considerably higher than realistic.  On the other hand, to understand the mathematical properties of the equation it may be
useful to go to much higher currents.

 An obvious approach to high current is to begin with the Gaussian and increase $I$ in steps, taking a solution at $I$ as the first guess for an attempted solution at $I+\Delta I$. An improvement to this idea can be achieved at little cost by instead using a linear extrapolation in  $I$:
\be
\vph(I+\Delta I)\approx \vph(I)+\frac{d\vph(I)}{dI}\Delta I\ . \label{extrap}
\ee
The derivative is found by differentiating the $I$-dependent equation with respect to $I$:
\bea
&&F(\vph(I),I)=0\ ,\nonumber \\
&& \sum_j\frac{\ptl F_i}{\ptl \vph_j}\frac{d\vph_j}{dI}+\frac{\ptl F_i}{\ptl I}=0\ . \label{diffI}
\eea
At the end of the Newton iteration for current $I$ we have in hand both the Jacobian $\ptl F(I)/ \ptl \vph$ and the quantity $\ptl F(I)/\ptl I$ (from the second term of (\ref{Feq0dis}). Thus it takes only one solution
of the linear system (\ref{diffI}) to produce the required  derivative $d\vph(I)/dI$ for (\ref{extrap}).

A convenient way to arrange the code is to make this method of advancing $I$ always available, and so that the case of a single $I$ is merely a special case.
Thus one specifies the initial and final values of $I$, and the number of intermediate values, taken to be evenly spaced. This is convenient for plotting $I$-dependent
quantities such as the centroid position or the r.m.s. bunch length of the Ha\"issinski distribution, and also for exploring the high current regime.

\section{Tests of the method for quasi-realistic wake potentials \label{test} } We consider examples of the  wake potential, obtained by solving Maxwell's equations with a quasi-realistic model
of the vacuum chamber providing the boundary conditions on metallic walls. The ideal wake potential $W_0(q)$, often called the delta wake, would be the
longitudinal field ${\mathcal E}(q,s)$ at a
fixed normalized distance $q$ from a point charge circulating on the ideal orbit, averaged  in the position $s$ over one turn. In practice the point charge is replaced
by a short Gaussian charge distribution ( a ``driving bunch") to provide the approximated wake potential $W(q)$. This smooth function could be called
 a ``pseudo - Green function" to distinguish it from a true Green function which cannot be smooth. If curvature of the orbit is neglected,
$W_0(q)$ displays ``causality", in that it vanishes in front of the point charge ($q>0$). In contrast $W(q)$ will be non-zero for small $q>0$, but will fall off quickly
with increasing $q$.

Pioneering simulations of wake potentials were carried out  for the damping rings of the Stanford Linear Collider (SLC), . The model was axially symmetric, with the fields being computed by Weiland's code TBCI \cite{tbci}. The boundary conditions were for infinite conductivity of the chamber walls. There were two calculations, one for the original vacuum chamber  \cite{bane1}, and one for a new vacuum chamber designed to have smoother walls \cite{bane2}.  The latter replaced the original in an attempt to gain a higher threshold in current for a bunch instability.

Computer power and codes for electromagnetics have been greatly improved since the work for the SLC, and the physical model for newer storage rings has
necessarily been extended to include coherent synchrotron radiation from curved orbits. A shorter driving bunch was needed, owing to shorter bunches in
the new rings, and better electromagnetic codes allowed the inclusion of three-dimensional structures. An example of this more modern effort is a calculation for
the low energy ring (LER) of KEKB (which is now out of service)~\cite{zhou}. The model  included CSR and resistive wall contributions as well as geometric wake fields.
This ring allowed a configuration with negative momentum compaction \cite{ikeda}, so we want to include that case in the Ha\" issinski solutions.

A rather different example of an ambitious calculation was for the positron ring of DAFNE  at Frascati \cite{zobov1, zobov2}. There CSR was not important
owing to the large bunch length, but the geometric structures were modeled very carefully.

For each of the examples mentioned we make a cubic spline interpolation of the wake potential data from the relevant simulation, then integrate the spline analytically
to make a smooth representation of the integrated potential $S(q)$ for input to the integral equation (\ref{ie}).

For a qualitative comparison of the various cases it is useful to see how much the collective force $-F$ resembles that from a linear combination
of purely inductive and purely resistive components. The corresponding wake potentials are $W(q)=a\delta\pr(q)$ (inductive) and $W(q)=-b\delta(q)$ (resistive),
where $a$ and $b$ are dimensionless positive constants. The corresponding force components are proportional to
\be
\int \delta\pr(q-q\pr)\lambda(q\pr)dq\pr=\lambda\pr(q)\ ,\quad  -\int \delta(q-q\pr)\lambda(q\pr)dq\pr=-\lambda(q)\ .  \label{indv}
\ee
Signs are determined by the requirement that there be energy loss from particles at the front of the bunch. We make a weighted least-squares fit
to the actual $F$ by minimizing the following integral with respect to $a$ and $b$:
\be
\int\lambda(q)\bigg[I\int W(q-q\pr)\lambda(q\pr)dq\pr -a\lambda\pr(q)+b\lambda(q)\bigg]^2dq\ .
\ee

Another way to make a qualitative comparison of cases is to plot the bunch centroid and the r.m.s. bunch length as a function of current. Such plots,
along with the fit to the inductive plus resistive wake, will be given for each of our examples.

\subsection{SLC damping ring with the original vacuum chamber}
For details of this example see Ref.~\cite{bane1}. For an RF voltage of 800 KeV the relevant parameters are as follows:
\be
\nu_s=0.0117\ ,\quad \sigma_E=0.805~{\rm MeV}\ ,\quad \sigma_z=4.95~{\rm mm}\ ,\quad I/N=2.71\cdot 10^{-12}~ {\rm pC/V}\ .
\ee
For a typical bunch population of $N=5\cdot 10^{10}$ the normalized current is $I=0.136~$ pC/V.
\begin{figure}[htb]
   \centering
   \includegraphics[width=1.1\linewidth]{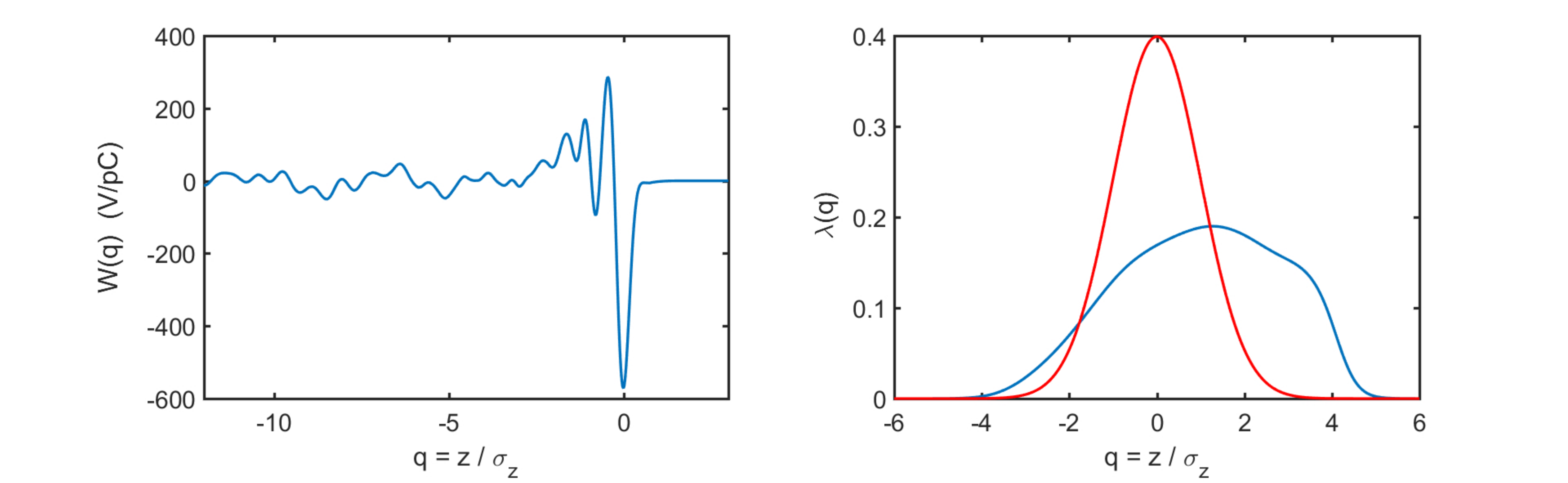}
  \caption{Results for the SLC damping ring with its original vacuum chamber. Left: Wake potential $W(q)$;
  Right: Equilibrium charge density for $N=5\cdot 10^{10}$ (blue) and in the limit of zero current (red).}
   \label{fig:wake_lamb_dr1}
\end{figure}
\begin{figure}[htb]
   \centering
   \includegraphics[width=1.1\linewidth]{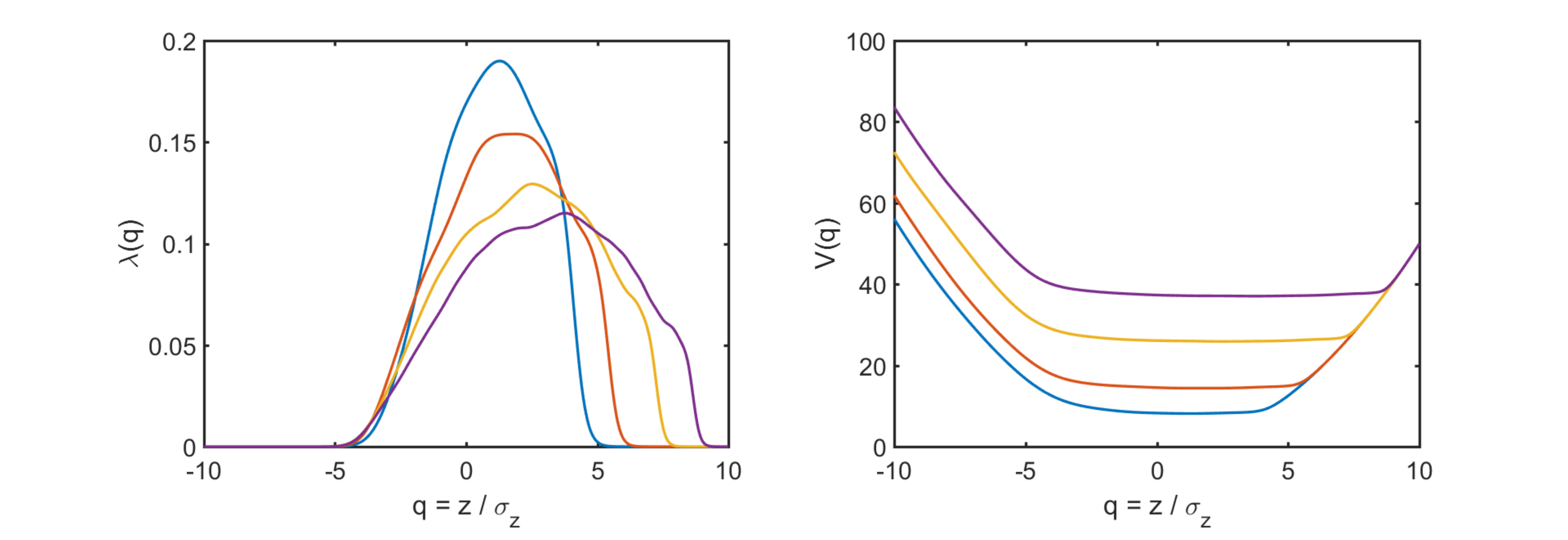}
  \caption{Ha\" issinski charge density for SLC damping ring with original vacuum chamber, for $N=(5,10,20,30)\cdot 10^{10}$ (left) and
  corresponding distorted potential well (right)}
   \label{fig:lambda_dr1}
\end{figure}
\begin{figure}[htb]
   \centering
   \includegraphics[width=1.1\linewidth]{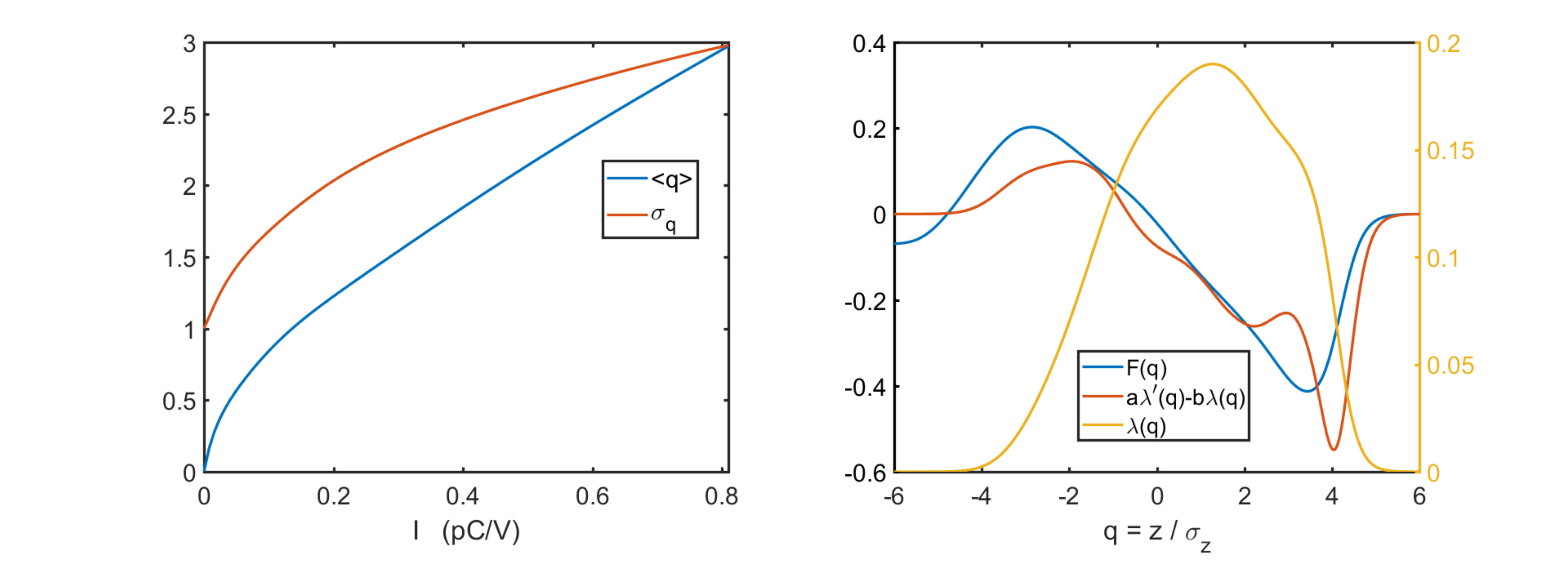}
  \caption{Left: Bunch centroid $<q>$ and r.m.s. length $\sigma_q$ as a function of normalized current, for SLC damping ring with original vacuum chamber.
  For $N=5\cdot 10^{10}$ the value of $I$ is $0.136~$ pC/V.
  Right: A fit to $F(q)$ by a linear combination of resistive and inductive terms, with $a=24.5$  and $b=7.07$, at $N=5\cdot 10^{10}$. }
   \label{fig:qbsig_ulr_dr1}
\end{figure}

The wake potential $W(q)$ computed with a Gaussian driving bunch with $\sigma=.5$~mm is shown in Fig.\ref{fig:wake_lamb_dr1} (left). The finite extent of the driving bunch accounts for
the potential not being zero at small positive $q$.

For  solution of the integral equation we choose the weights $w_i$ for numerical quadrature to be those for Simpson's method; namely $\{w_i\}=(\Delta q/3)
(1,4,2,\cdots,2,4,1).$.  In the present and following
examples we define the mesh in (\ref{mesh}) with $\kappa=6$ for a mesh extending to $6 \sigma_z$, and $m=400$ for 801 mesh points.  We take $r=10^{-14}$ in (\ref{rdef}) as the criterion for convergence.  For a bunch population of $N=5\cdot 10^{10}$, roughly the maximum that was stored in
the ring, we get the Ha\"issinski solution shown in Fig.\ref{fig:wake_lamb_dr1} (right). For comparison we show the Gaussian solution for zero current.
The iteration to achieve this solution, beginning with the Gaussian, converged in 11 steps.

The good convergence is found to persist at much higher currents. In Fig.\ref{fig:lambda_dr1} (left) we compare solutions for $N=(5,10,20,30)\cdot 10^{10}$. These solutions all started
with the Gaussian as first guess, but the extent of the mesh had to be increased (taking $\kappa=10$) because of the increased bunch lengthening. At the highest current,
39 iterations were required. It is interesting to find that the value of $r$ at the first iterate is always rather large, say 0.25, even at very small current.
The same sequence of solutions is obtained by applying the method of continuation in $I$ presented in Section \ref{newton}. After a small step in $I$ very few
iterations are needed for convergence.

The very pronounced bunch lengthening in this example corresponds to a flat bottom in the distorted potential well. The well as given by (\ref{vwiths}) is shown
in Fig.(\ref{fig:lambda_dr1}) (right), for the same sequence of currents. Since $\log\lambda(q)=-V(q)-\log A$, the wavy modulations in $\lambda$ at high current must have a
counterpart in $V(q)$. Taking the logarithm makes the modulations  too small to be apparent on the scale of the graph of $V(q)$.

The fit to a sum of purely inductive and resistive wakes, plotted in Fig.\ref{fig:qbsig_ulr_dr1} (right), shows that the inductive character is dominant within the
bunch distribution: $a/b=3.47$. A purely inductive wake lengthens the bunch while keeping it symmetric about $q=0$,  while a purely resistive wake makes the bunch lean forward with little change in its length. Accordingly, the bunch form overlayed in Fig.\ref{fig:qbsig_ulr_dr1} shows relatively little leaning.

In Fig.\ref{fig:qbsig_ulr_dr1} (left) we show the evolution with current of the normalized bunch length and centroid position. For $N=5\cdot 10^{10}$ the normalized
current is $I=0.136~$ pC/V. Both $\sigma_q$ and $<q>$ show a steady increase, almost linear at high $I$.

The computations were done by a  Fortran code using standard software for the linear algebra to solve for Newton iterates. The latter provides an estimate of the
condition number of the Jacobian matrix, which turned out to be acceptably small, ranging from 10 at nominal current to 86 at six times nominal. Thus the iterates are numerically well-defined.  The computation time was negligible. In the following examples, the convergence and condition numbers were no worse than in the present
case, and often better.

\subsection{SLC damping ring with the improved vacuum chamber}
This case is reviewed in Refs.\cite{bane2, bane3}. For an RF voltage of 800 KeV the relevant parameters are the following:
\be
\nu_s=0.0116\ ,\quad \sigma_E=0.847~{\rm MeV}\ ,\quad \sigma_z=4.95~{\rm mm}\ ,\quad I/N=2.60\cdot 10^{-12}~ {\rm pC/V}\ .
\ee
For a typical bunch population of $5\cdot 10^{10}$ we have $I=0.14$ pC/V.

\begin{figure}[htbp]
   \centering
    \includegraphics[width=1.1\linewidth]{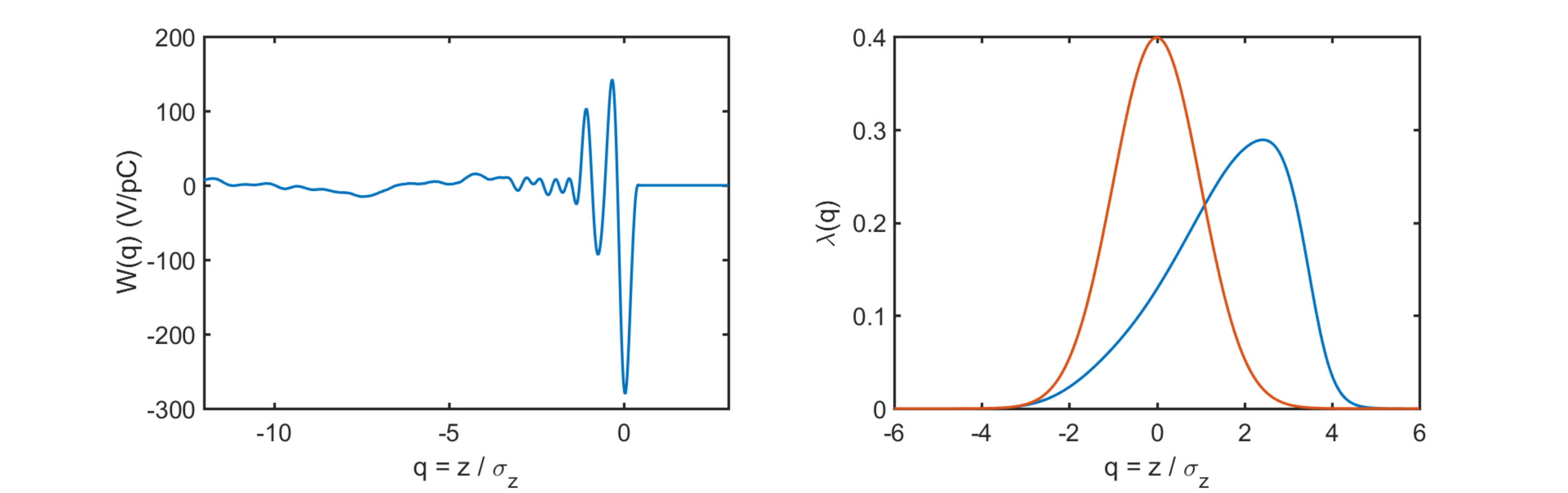}
  \caption{Results for the SLC damping ring with improved vacuum chamber. Left: Wake potential $W(q)$;
  Right: Equilibrium charge density for $N=5\cdot 10^{10}$ (blue) and in the limit of zero current (red).}
   \label{fig:wake_lamb_dr2}
\end{figure}

\begin{figure}[htbp]
   \centering
 \includegraphics[width=1.1\linewidth]{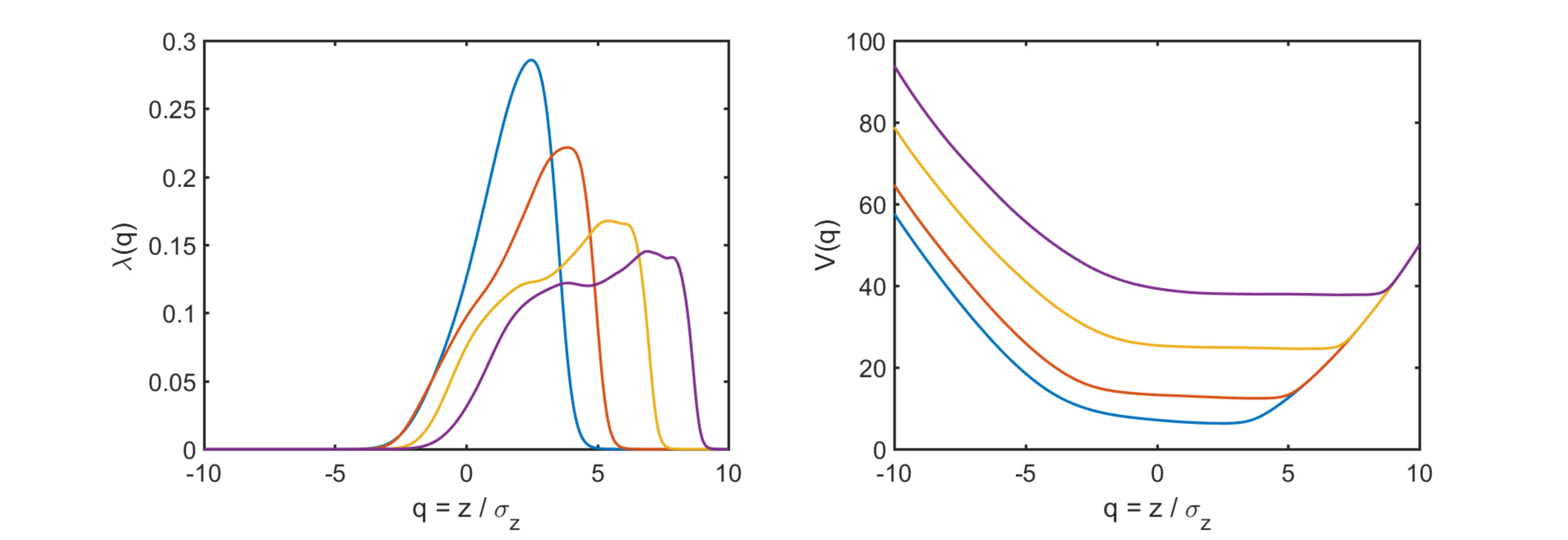}

  \caption{ Ha\" issinski charge density for SLC damping ring with improved vacuum chamber, for $N=(5,10,20,30)\cdot 10^{10}$ (left) and
  corresponding distorted potential well (right)}
   \label{fig:lamb_well_dr2}
\end{figure}

\begin{figure}[htbp]
   \centering
    \includegraphics[width=1.1\linewidth]{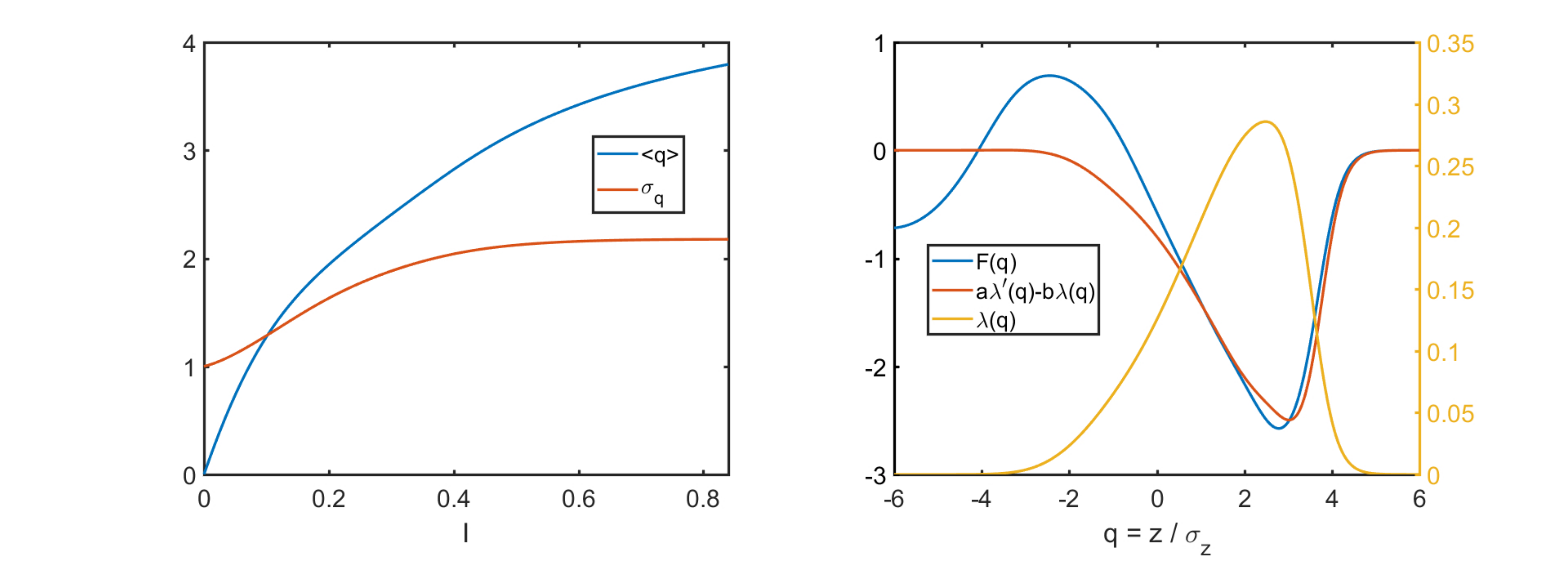}
  \caption{Left: Bunch centroid $<q>$ and r.m.s. length $\sigma_q$ as a function of normalized current, for SLC damping ring with improved vacuum chamber.
  For $N=5\cdot 10^{10}$ the value of $I$ is $0.140~$ pC/V.
  Right: A fit to $F(q)$ by a linear combination of resistive and inductive terms, with $a=3.06$  and $b=8.14$, at $N=5\cdot 10^{10}$.}
   \label{fig:qbarsig_dr2}
\end{figure}

In Figs.~\ref{fig:wake_lamb_dr2} - \ref{fig:qbarsig_dr2} we see a marked change in comparison to the case of the original damping ring.
 In Fig.\ref{fig:qbarsig_dr2} (right) the resistive component is shown to dominate the inductive: $b/a= 2.66$. The bunch
leans forward in the RF bucket, to compensate for the energy loss from the resistive wake field. The bunch length tends to saturate with
increasing current, as is seen in Fig.\ref{fig:qbarsig_dr2} (left). The fall-off of charge density at the leading edge becomes sharper and sharper as the current increases, as is seen in Fig.\ref{fig:lamb_well_dr2}.


\clearpage
\subsection{KEKB Low Energy Ring}
This case is reviewed in Refs.\cite{zhou,ikeda}. For an RF voltage of 800 KeV the relevant parameters are the following:
\be
\nu_s=0.024\ ,\quad \sigma_E=2.54~{\rm MeV}\ ,\quad \sigma_z=4.58~{\rm mm}\ ,\quad I/N=4.18\cdot 10^{-12}~ {\rm pC/V}\ .
\ee
For a typical bunch population of $6.6\cdot 10^{10}$ we have $I=0.0275$ pC/V.

\begin{figure}[htb]
   \centering
    \includegraphics[width=1.1\linewidth]{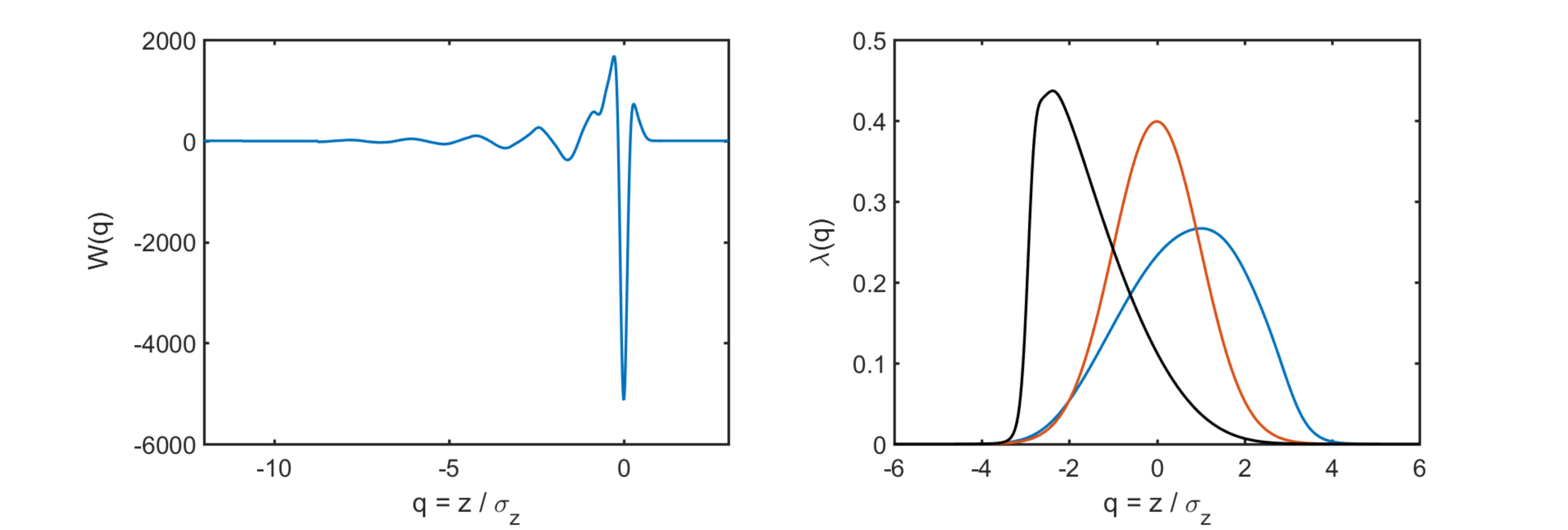}
  \caption{Results for KEKB-LER. Left: Wake potential $W(q)$;
  Right: Equilibrium charge density for $N=6.6\cdot 10^{10}$ (blue), in the limit of zero current (red),
  and for negative momentum compaction (black)}
   \label{fig:wake_lamb_kek}
\end{figure}

This ring was able to run with negative momentum compaction. In Fig.\ref{fig:wake_lamb_kek} (right) we plot a charge density for that case in the black
curve. This was obtained by changing the sign of $I$, while keeping all other parameters unchanged; see (\ref{Idef}). The steep fall-off at the back of the
bunch is typical for negative momentum compaction.

\begin{figure}[htb]
   \centering

    \includegraphics[width=1.1\linewidth]{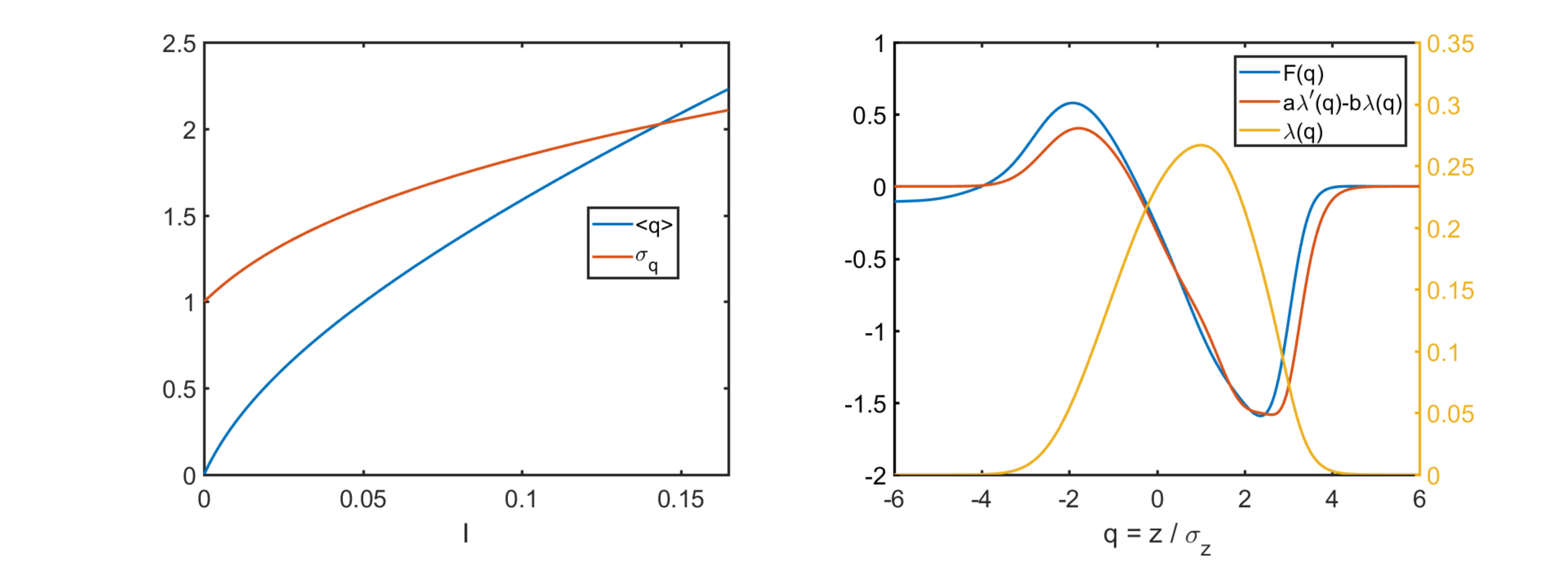}
  \caption{Left: Bunch centroid $<q>$ and r.m.s. length $\sigma_q$ as a function of normalized current, for KEKB-LER.
  At $N=6.6\cdot 10^{10}$ the value of I is $0.0275~$ pC/V.
  Right: A fit to $F(q)$ by a linear combination of resistive and inductive terms, with $a=7.45$  and $b=3.41$, at $N=6.6\cdot 10^{10}$.}
   \label{fig:qbarsig_kek}
\end{figure}

In this example the linear combination of inductive and resistive components provides a remarkably accurate representation of the collective force,
as is seen in Fig.\ref{fig:qbarsig_kek} (right). The inductive part dominates moderately, with $a/b=2.18$, so that we see the inductive pattern
of strong bunch lengthening with relatively little forward tipping as the current is increased. Compare the behavior of bunch length versus current
in Fig.\ref{fig:qbarsig_kek} (left) with the corresponding graph Fig.\ref{fig:qbarsig_dr2} (left) for the previous example, in which the resistive part dominated.

\begin{figure}[htb]
   \centering

   \includegraphics[width=1.1\linewidth]{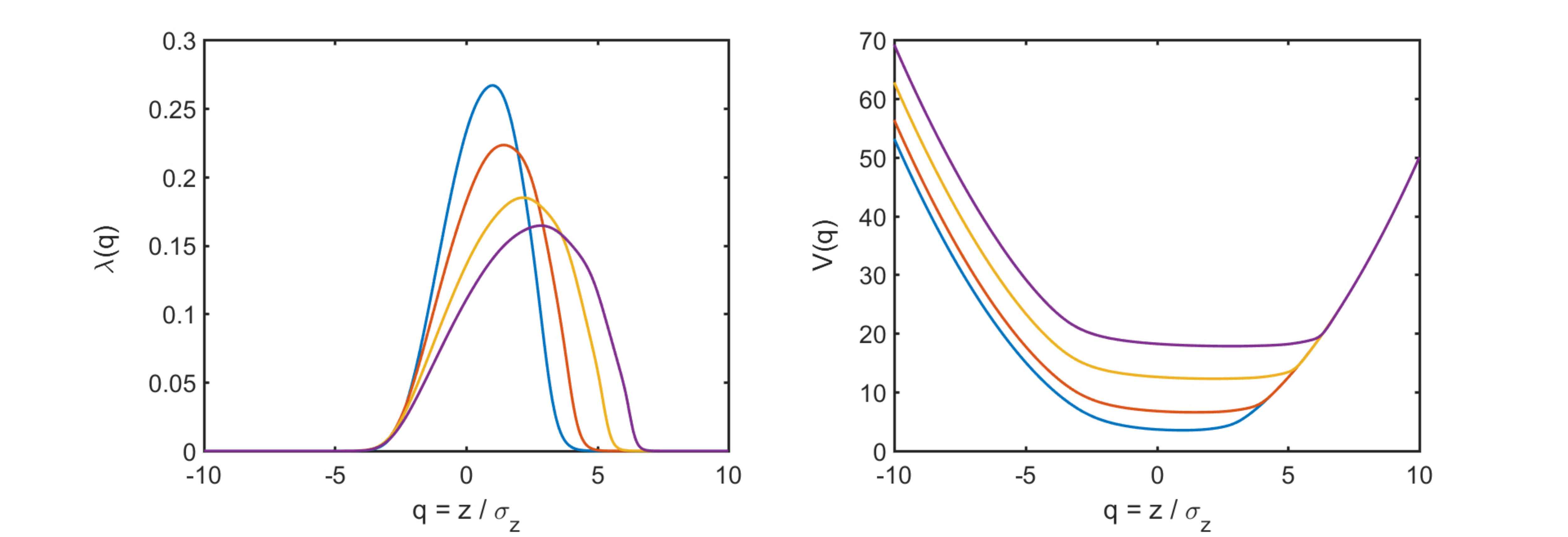}
  \caption{Ha\" issinski charge density for KEKB-LER, for $N=(6.6,13.2,26.4,39.6)\cdot 10^{10}$ (left) and
  corresponding distorted potential well (right)}
   \label{fig:lamb_well_kek}
\end{figure}
\clearpage
\subsection{DAFNE Positron Ring}
See Refs.\cite{zobov1,zobov2} for information on this case. For an RF voltage of 250 kV we have
\be
 \nu_s=0.011\ ,\quad \sigma_E=0.202~{\rm MeV}\ ,\quad \sigma_z=2~{\rm cm}\ ,\quad I/N=1.15\cdot 10^{-11}~ {\rm pC/V}\ .
\ee
For a typical bunch population of $N=9\cdot 10^{10}$ we have $I=1.035$ pC/V.

\begin{figure}[htb]
   \centering
    \includegraphics[width=\linewidth]{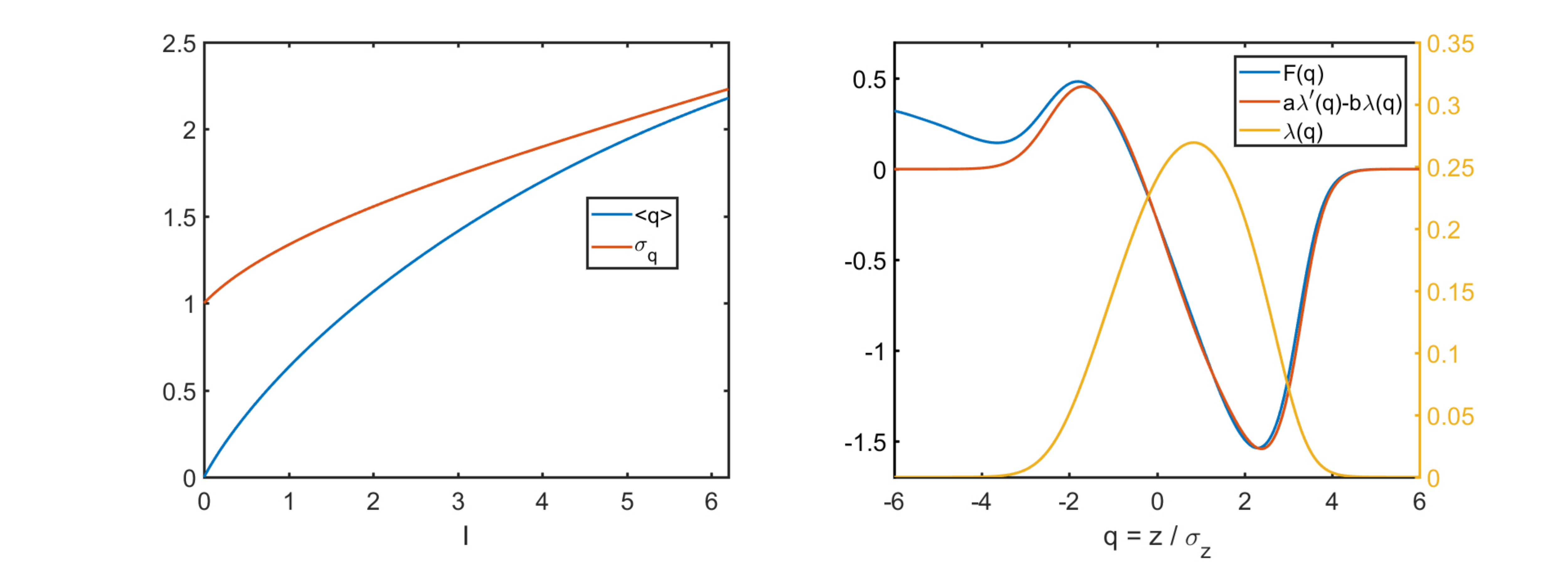}
  \caption{Left: Bunch centroid $<q>$ and r.m.s. length $\sigma_q$ as a function of normalized current, for DAFNE.
  At $N=9\cdot 10^{10}$ the value of $I$ is $1.035~$pC/V.
  Right: A fit to $F(q)$ by a linear combination of resistive and inductive terms, with $a=7.45$  and $b=3.21$, at $N=9\cdot 10^{10}$.}
   \label{fig:qbarsig_dafne}
\end{figure}

Although the parameters of this ring are totally different from those of KEKB, especially in the long  bunch length, the qualitative picture of wakes and bunch forms
is remarkably similar in our normalized variables.  Again we have a very good fit to a sum of inductive and resistive components, with almost a 2:1 ratio of inductive to
resistive parts, as is seen in Fig.\ref{fig:qbarsig_dafne} (right). The pattern of bunch forms and bunch length versus current is very similar to that of KEKB.
\begin{figure}[htb]
   \centering
    \includegraphics[width=1.1\linewidth]{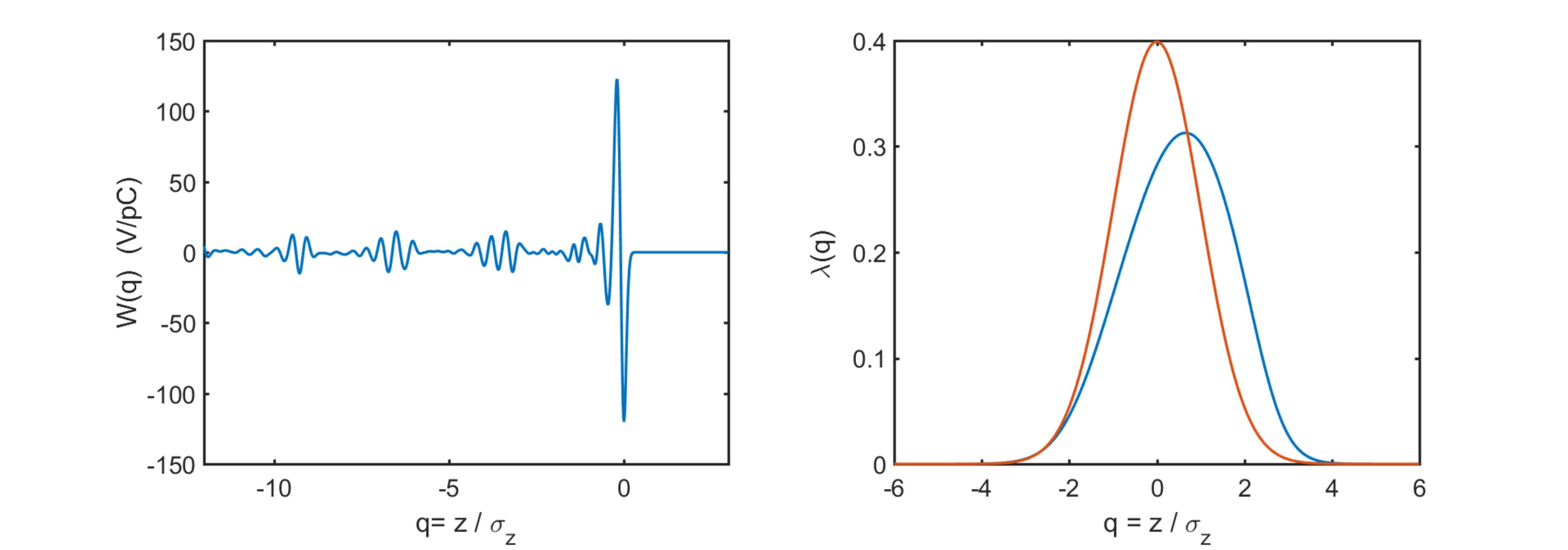}
  \caption{Results for DAFNE positron ring. Left: Wake potential $W(q)$;
  Right: Equilibrium charge density for $N=9\cdot 10^{10}$ (blue), and in the limit of zero current (red)}
   \label{fig:wake_lamb_dafne}
\end{figure}
\begin{figure}[htb]
   \centering
   \includegraphics[width=1.1\linewidth]{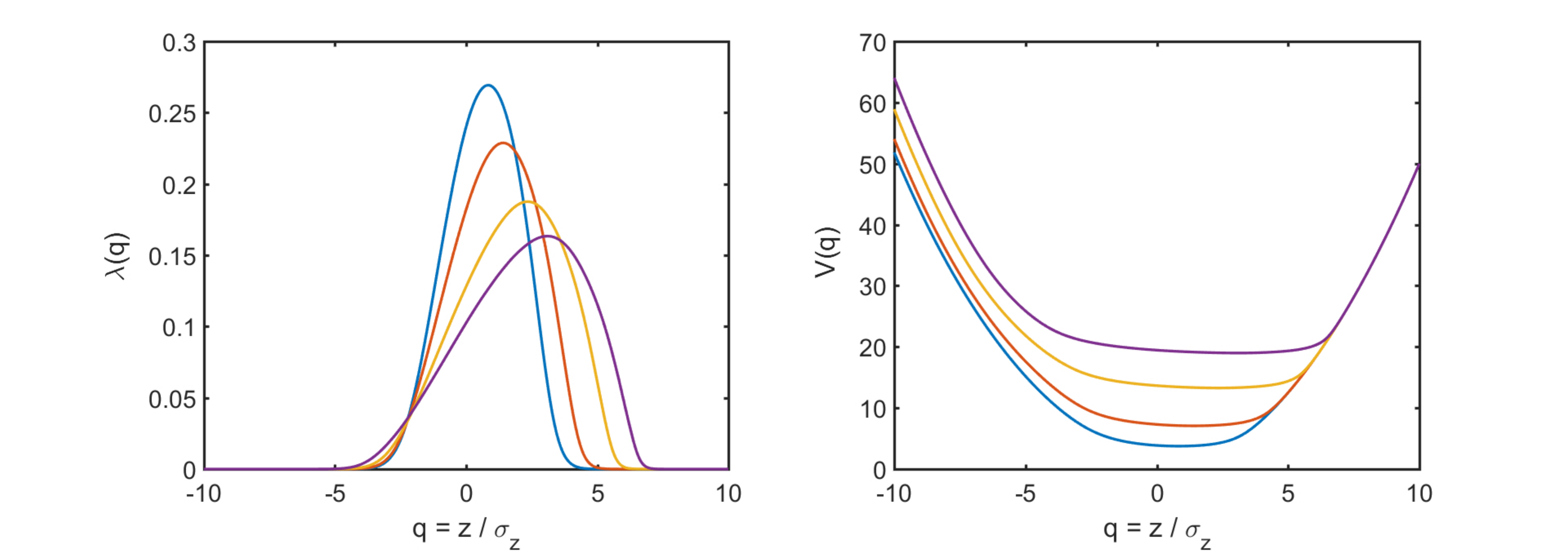}
  \caption{Ha\" issinski charge density for DAFNE, for $N=(9,18,36,54)\cdot 10^{10}$ (left) and
  corresponding distorted potential well (right)}
   \label{fig:lamb_well_dafne}
\end{figure}
\clearpage
\section{Conclusions and Outlook}
We have demonstrated a simple and convenient method to solve the Ha\" issinski equation, with quasi-realistic wake potentials for different kinds of electron storage rings. In
work not covered in this report, we have also verified that the method works as well
for the broad band resonator model of the wake potential, with similar or better experience regarding convergence.
We have also applied the method with the wake from coherent synchrotron radiation, accounting for the ``shielding" due to
the vacuum chamber. The parallel plate model of the vacuum chamber was invoked in \cite{bcs}, and the toroidal model with resistive wall in \cite{marit}.

There is scope for  mathematical analysis of the Ha\" issinski equation, which we hope to present in a later paper.
One can prove existence and uniqueness of solutions at sufficiently small current, under weak conditions on the wake potential. Also, a critique of previous work on ideal models of the wake potential seems to be in order. The models of
purely inductive, purely resistive, and purely capacitive wake potentials involve some interesting mathematical questions that should be re-examined.

\section{Acknowledgments}
We greatly appreciate the help of Demin Zhou and Mikhail Zobov, who  provided files of the wake potentials
for KEKB and DAFNE, respectively. Our work was supported in part by the U.S. Department of Energy, Office of Science,
Program in High Energy Physics, under Award No. DE-AC03-76SF00515.

\end{document}